\documentclass[twocolumn]{aastex631}

\shorttitle{Meridional Circulation of Dust}
\shortauthors{Szul\'agyi et al.}

\graphicspath{{./}{figures/}}

\begin{document}

\title{Meridional Circulation of Dust and Gas in the Circumstellar Disk: \\
Delivery of Solids onto the Circumplanetary Region}

\author[0000-0001-8442-4043]{J. Szul\'agyi}
\affiliation{Institute for Particle Physics and Astrophysics, ETH Z\"urich, Switzerland}
\affiliation{Center for Theoretical Astrophysics and Cosmology, Institute for Computational Science, University of Z\"urich,\\ Winterthurerstrasse 190, CH-8057 Z\"urich, Switzerland}

\author{F. Binkert}
\affiliation{University Observatory, Faculty of Physics, Ludwig-Maximilians-Universit\"t M\"unchen,
Germany}
\affiliation{Institute for Particle Physics and Astrophysics, ETH Z\"urich, Switzerland}

\author{C. Surville}
\affiliation{Center for Theoretical Astrophysics and Cosmology, Institute for Computational Science, University of Z\"urich,\\ Winterthurerstrasse 190, CH-8057 Z\"urich, Switzerland}



\begin{abstract}
We carried out 3D dust+gas radiative hydrodynamic simulations of forming planets. We investigated a parameter grid of Neptune-, Saturn-, Jupiter-, and 5 Jupiter-mass planets at 5.2, 30, 50 AU distance from their star. We found that the meridional circulation \citep{Szulagyi14,FC16} drives a strong vertical flow for the dust as well, hence the dust is not settled in the midplane, even for mm-sized grains. The meridional circulation will deliver dust and gas vertically onto the circumplanetary region, efficiently bridging over the gap. The Hill-sphere accretion rates for the dust are $\sim10^{-8}$ to $10^{-10}$ $\rm{M_{Jup}/yr}$, increasing with planet-mass. For the gas component, the gain is $10^{-6}$ to $10^{-8}$ $\rm{M_{Jup}/yr}$. The difference between the dust and gas accretion rates is smaller with decreasing planetary mass. In the vicinity of the planet, the mm-grains can get trapped easier than the gas, which means the circumplanetary disk might be enriched with solids in comparison to the circumstellar disk. We calculated the local dust-to-gas ratio (DTG) everywhere in the circumstellar disk and identified the altitude above the midplane where the DTG is 1, 0.1, 0.01, 0.001. The larger the planetary mass, the higher the mm-sized dust is delivered and a larger fraction of the dust disk is lifted by the planet. The stirring of mm-dust is negligible for Neptune-mass planets or below, but significant above Saturn-mass. 
\end{abstract}



\section{Introduction} \label{sec:intro}

Dust processes in circumstellar disks are extremely crucial to understanding disk evolution, the formation of planetary building blocks, chemical processes in the disks and the observability of these disks.

Observations of circumstellar disks provide an insight into the dust settling and dust-to-gas ratios of these disks. Recently, several edge-on disk studies were made to better understand the dust settling and how it is affected by various mechanisms, e.g. turbulence \citep[e.g.][]{Kasper15,Villenave20,Wolff21}. The dust-to-gas ratio of the circumstellar disk is crucial for multiple processes, for example the formation of planetesimals via streaming instability \citep[e.g.][]{Goodman00,Youdin05}, that requires a local dust-to-gas ratio of one. Observations nowadays can estimate the (global) dust-to-gas ratio in disks and compare them to theoretical model predictions \citep{Turrini19,Horne12}. Hence, to understand which part of the disk can have a high enough dust-to-gas ratio for streaming instability to operate, we have to rely on dust+gas hydrodynamic simulations.

So far, dust+gas hydrodynamic simulations of forming planets have been done in 2D, mainly concentrating on studying the gap(s) that the planets opened, the vortices and other disk sub-structures \citep[e.g.][]{Baruteau16,Zhu16,Surville16,Jin16,Fedele17,Tamfal18,Dong18,Zhang18,Baruteau19,Surville19,Ricci18,Pinte19,WF20}. In the past, 3D grid-based dust-gas hydrodynamical simulations {studied} planet-less processes, e.g. streaming instability with local box simulations, or magneto-hydrodynamic simulations where magnetic effects {influence} the dust \citep[e.g.][]{Johansen2007,Chiang09,Flock20,Krapp19,Paardekooper20,Paardekooper21,McNally21,Zhu21,Lin21}. Studies including planets were only done with solids (larger dust grain sizes, planetesimals) treated as particles. There is a need for grid-based, multifluid 3D dust-gas hydrodynamical simulations with embedded planets to understand how the dust {flows and processes} change in this case. The multifluid approach can really simulate the (fine) dust, and it inherently includes the dust feedback on the gas, an important process that many of the particle works omit \citep{Fu14,Kanagawa18}.

In recent years, there have been investigations about the gas flows in circumstellar disks. In particular, the meridional circulation of gas \citep{Szulagyi14,FC16} showed that the spiral wakes of the planet are continuously reprocessing the circumstellar disk material. They bring up material from the midplane, deposit the gas in the higher parts of the disk. The gas then flows into the gap and onto the circumplanetary region. The part of the gas which is not immediately accreted by the inner circumplanetary disk regions, will leave the  circumplanetary disk region \citep{Szulagyi14,FC16,Teague19}, and go back to the circumstellar disk through the spiral wakes of the planet, then rises up again to maintain this vertical (meridional) motion \citep{Szulagyi14,Morbidelli14}. Not only the massive planets can stir up the gas in their vicinity. In the case of terrestrial mass planets, the same circulation was found, although on smaller scales. In the rocky planet mass regime the process is called atmospheric recycling \citep{Ormel15,Ormel15b,Cimerman17} and the circulation is more concentrated on the planet vicinity and it is forming the first atmospheres of these terrestrial planets. The two processes (meridional circulation and atmospheric recycling) have important similarities: the planet {accretes} vertically from the polar regions, and there is an outflow in the midplane. The meridional circulation does not only exist in computer simulations, but just recently has been directly observed with ALMA \citet{Teague19}. This shows the importance to study further the process and understand how the dust is influenced by the meridional circulation.

In the past, it was generically accepted that dust settles toward the midplane \citep{DD04,Nomura06,Krijt16}. Only the small dust grains (micron-sized or so) that are light enough to follow the gas motion in the disk can be found near the surface of the disk. Millimeter-sized and larger grains are expected to be found only in the midplane \citep{DD04,Nomura06,Krijt16}. It was however never studied before how the dust is influenced by the meridional circulation and the spiral wakes of the planet. In our previous paper \citep{Binkert21} we found that the dust is stirred up vertically by the planets, and the dust disk is therefore very puffed up in the vertical direction, even for mm-sized grains. In this paper, we will investigate the kinematics of the problem, study the flow of the meridional circulation for the dust, the delivery of dust onto the circumplanetary disk, as well as the local dust-to-gas ratio in the circumstellar disk \citep{Horne12} that is influenced by the dust stirring.

\section[]{Methods}
\label{sec:numerical}

\subsection{Hydrodynamic Simulations}
\label{sec:hydro}

We have carried out 3D radiative thermo-hydrodynamic simulations of gas and dust with the JUPITER code \citep{Szulagyi16,Binkert21} that was developed by F. Masset and J. Szul\'agyi. Apart from solving for the compressible Navier-Stokes equations, the code also solves radiative transfer in the flux-limited diffusion approximation with the two-temperature approach \citep{Commercon11}. Furthermore, the dust component is treated via a multi-fluid approach \citep{Nakagawa86} with the details described in \citet{Binkert21}. The coordinate system was Spherical, centered on the Solar-mass star that was treated as a point-mass with an effective temperature of 5700\,K. Stellar irradiation from the Solar-analog star was included. We carried out a parameter study, where we considered four different planetary masses and three semi-major axes, which totals up to 12 simulations. The different planetary masses were a Neptune, a Saturn, a Jupiter and a 5 Jupiter-mass planet placed at 5.2\,AU, 30\,AU and 50\,AU from their star. The circumstellar disk initial surface density value at the planet's location, the inner and outer disk limits were scaled accordingly to the semi-major axes, their values are listed in Table \ref{tab:params}. The kinematic viscosity $\nu$ was constant throughout the disk with a value of $10^{15}\:\mathrm{cm^2/s}$ which corresponds to a turbulent alpha parameter of $4.7\cdot10^{-3}$, $1.9\cdot10^{-3}$ and $9.8\cdot10^{-4}$ at 5.2 AU, 30 AU and 50 AU respectively.
The opening angle of the circumstellar disk was initially 7.4 degrees with a flaring index of 0.28. The initial surface density was constant throughout the disk, the slope was set to zero. The disk of course evolves from the initial set values, due to the heating/cooling and the perturbation of the planet. Before we added the planets into the simulations, we run the calculation for 150 orbits to reach thermal equilibrium of the circumstellar disk with all the heating and cooling mechanisms (stellar irradiation, adiabatic compression and expansion, viscous heating, radiative dissipation, shock-heating). Only after that we added the planet, which we grow slowly to its final mass over 100 orbits. We run the simulations with another 100 orbits, so altogether 200 orbits with the planet present.

\begin{table*}
\begin{center}
\begin{tabular}{l c c c c c c} 
 \hline
   name & $R_p$ [AU] & $M_p$ [$M_{Jup}$] & $\Sigma_{0,g}$ [g/cm$^2$]  & $\nu$ [cm$^2$/s] & $R_{in}$ [AU] &$R_{out}$ [AU] \\ 
  \hline
 m5au1nep &  5.2  & 0.05 & 120 & $1.016\cdot10^{15}$& 2.08   &   12.40\\ 
 m5au1sat &  5.2  & 0.3  & 120& $1.016\cdot10^{15}$ & 2.08   &   12.40\\ 
 m5au1jup &  5.2  & 1  & 120 &$1.016\cdot10^{15}$ & 2.08   &   12.40\\ 
 m5au5jup &  5.2  & 5  & 120 & $1.016\cdot10^{15}$& 2.08   &   12.40\\ 
 m30au1nep &  30  & 0.05  & 10.1 & $1.016\cdot10^{15}$ & 12.00  &    71.54 \\ 
 m30au1sat &  30  & 0.3  & 10.1  & $1.016\cdot10^{15}$ & 12.00  &    71.54 \\ 
{m30au1jup} &  30  & 1.0  & 10.1  & $1.016\cdot10^{15}$ & 12.00  &    71.54\\ 
 m30au5jup &  30  & 5.0  & 10.1 & $1.016\cdot10^{15}$ & 12.00  &    71.54 \\ 
 m50au1nep &  50 & 0.05  & 7.05 &$1.016\cdot10^{15}$ & 20.00 &    119.23\\
 m50au1sat &  50  & 0.3  & 7.05 &$1.016\cdot10^{15}$ &  20.00 &    119.23\\
 m50au1jup &  50  & 1.0  & 7.05 &$1.016\cdot10^{15}$ &  20.00 &    119.23\\
 m50au5jup &  50  & 5.0  & 7.05 &$1.016\cdot10^{15}$ &  20.00 &    119.23\\
 \hline
\end{tabular}
\caption{This table provides an overview of the 12 radiative hydrodynamic simulations which we carried out.}
\label{tab:params}
\end{center}
\end{table*}

 The adiabatic index was 1.43, while the mean molecular weight was set to 2.3. The heating processes included adiabatic compression, viscous heating and stellar irradiation, while the cooling processes were adiabatic expansion and radiative diffusion. The resolution was 680 cells azimuthally over $2\pi$, 215 cells radially, 20 cells in the co-latitude direction for the lower half of the disk. We assumed symmetry to the midplane, hence only the lower half of the circumstellar disk was simulated in each case to save computational time. The dust size was set to be 1 mm, which translates to a semi-coupled case with these parameters above. We chose 1 mm, because it is perhaps the most relevant for ALMA dust continuum, as well as pebble accretion \citep{Ormel10,Lambrechts12}. A smaller grain size, e.g. micron-sized dust would be well coupled to the gas, so  the {distribution of those dust particles is expected} to follow that of the gas. The opacity table used in these simulations was equivalent to what was used in \citet{Szulagyi19}, and included both gas and dust opacities. The initial global dust-to-gas ratio was 1\%, which then evolved throughout the simulation. The dust composition in the opacity calculation was assumed to be 40\% silicate, {40\% water-ice and 20\% amorphous carbons}, while gas component followed the one used in \citet{BL94} for the gases. These set of simulations somewhat differ from {\citet{Binkert21}}, for example, in the boundary condition setup: here we let dust to flow in from the outer radial boundary. In both papers we used a boundary condition in the vertical direction above the disk that is a Gaussian extrapolation of the disk density, but there is no inflow or outflow permitted. For the temperature field, we fixed the temperature of the ghost cells above the disk surface to 3K, i.e. the Cosmic Microwave Background temperature, to allow cooling to happen through the disk surface. Azimuthally, we had a periodic boundary condition, since we simulated the entire disk. 

\section{Results}

\subsection{Spiral Wakes of the Planet}
\label{sec:spiral}

\begin{figure*}
\includegraphics[width=18cm]{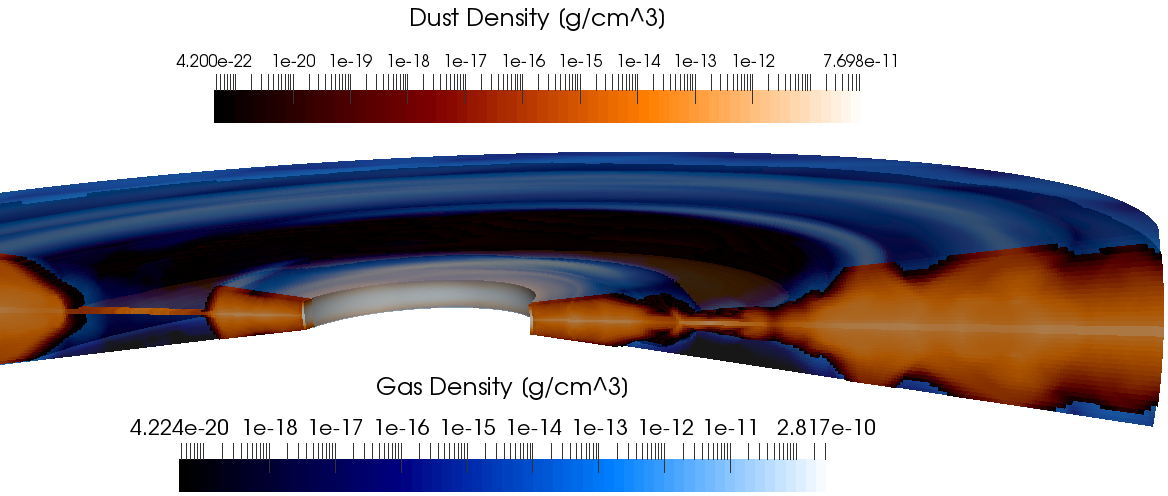}
\caption{Composite image of the gas (blue) and dust (orange) distribution in the circumstellar disk. The cut through the disk is near the planet's position. This simulation is the 5 Jupiter-mass case at 5.2 AU. Clearly, the dust distribution vertically is very extended. This is due to the meridional circulation, where the planet spiral arms stir up the disk material. The higher is the planet mass, the stronger is this stirring {(the dust flies higher, and radially farther from the planet)}.}
\label{fig:dustvertical}
\end{figure*}

In \citet{Binkert21}, we examined the vertical extension of the dust disk, due to the presence of the planet. There, we found the planet to continuously stir up the dust, hence in disks where massive planets (above Neptune-mass, see in \citealt{Binkert21}) are orbiting, moderately coupled dust does not settle completely toward the midplane (Fig. \ref{fig:dustvertical}).

\begin{figure*}
\includegraphics[width=9cm]{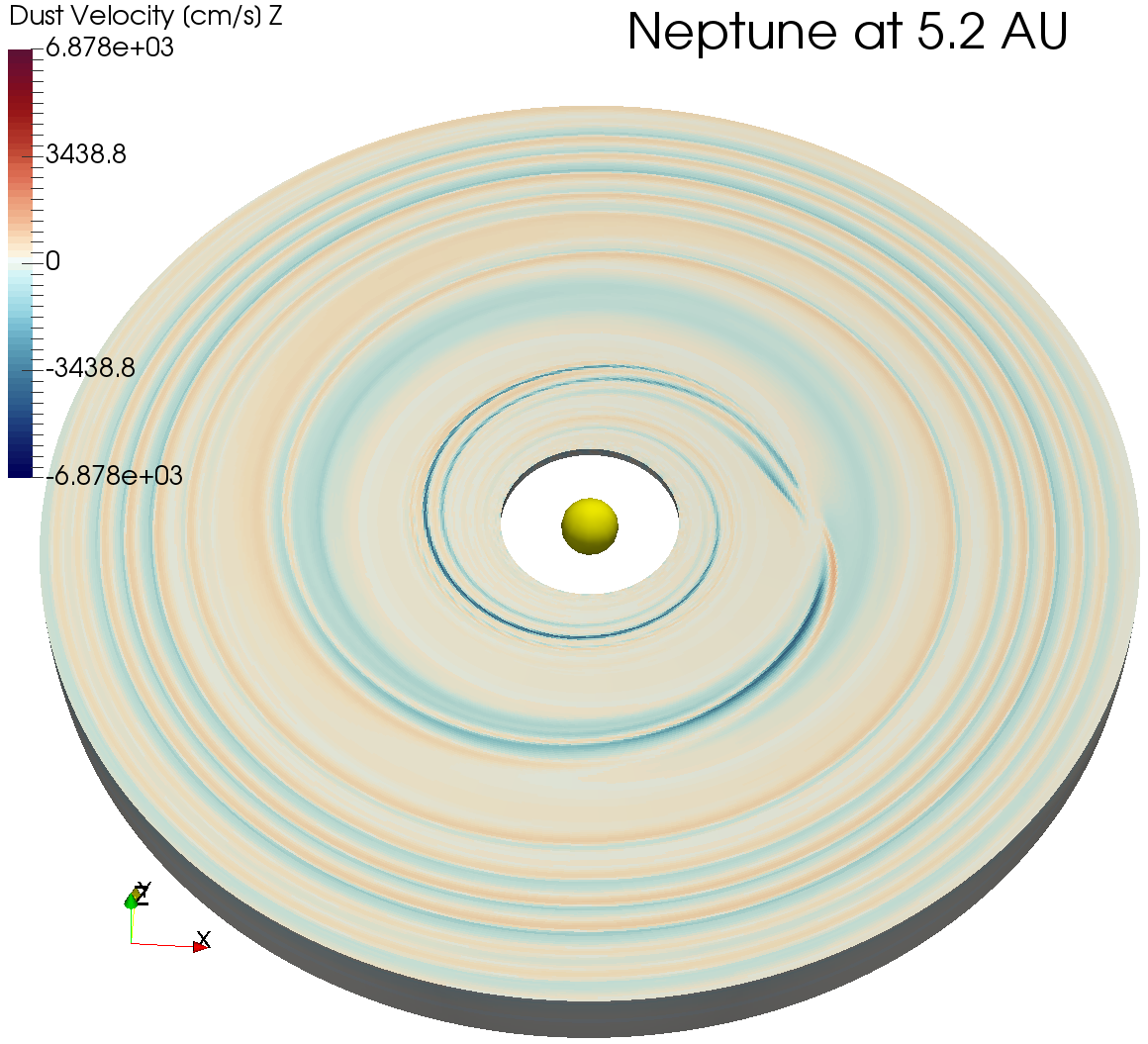}\includegraphics[width=9cm]{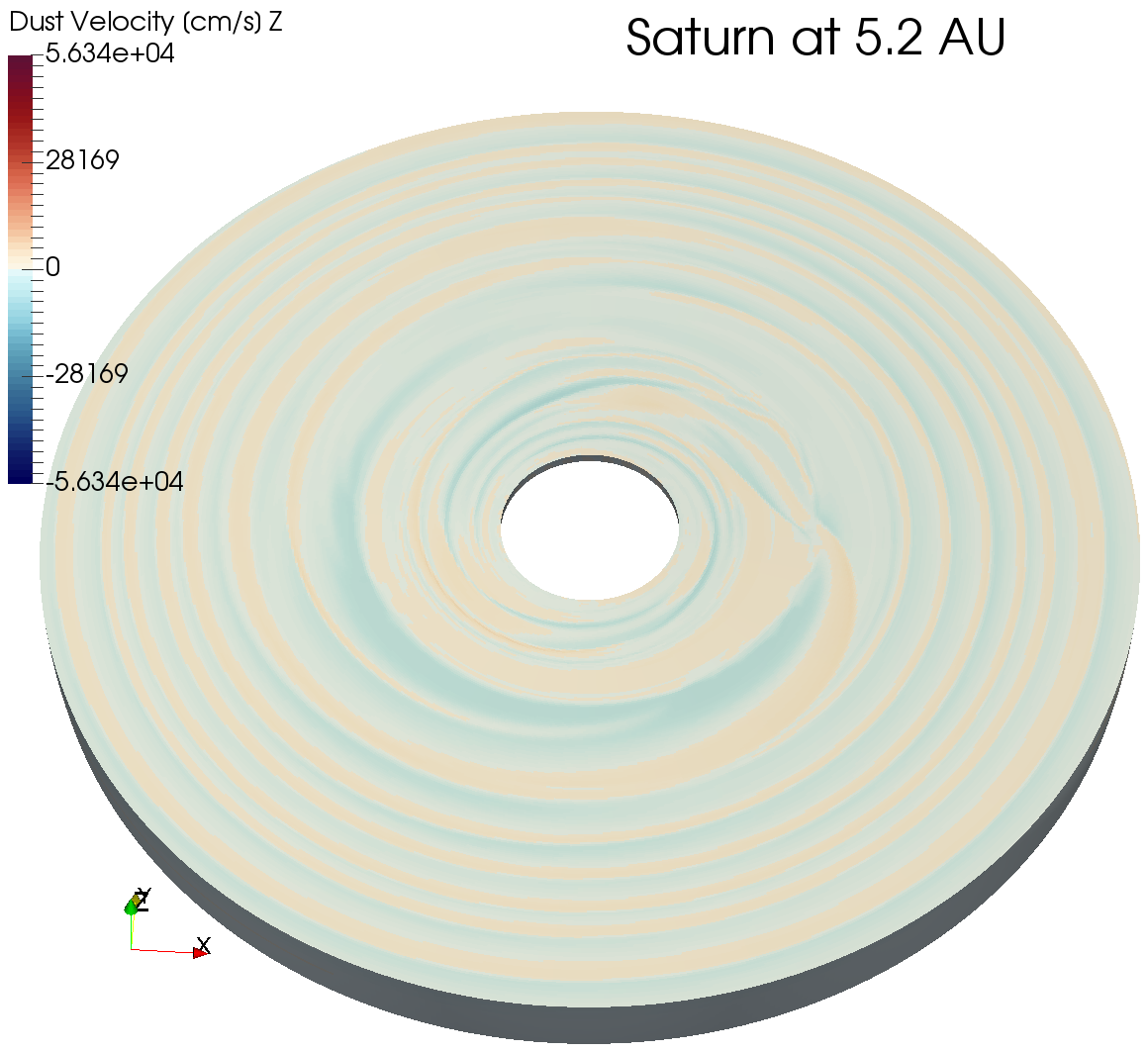}
\includegraphics[width=9cm]{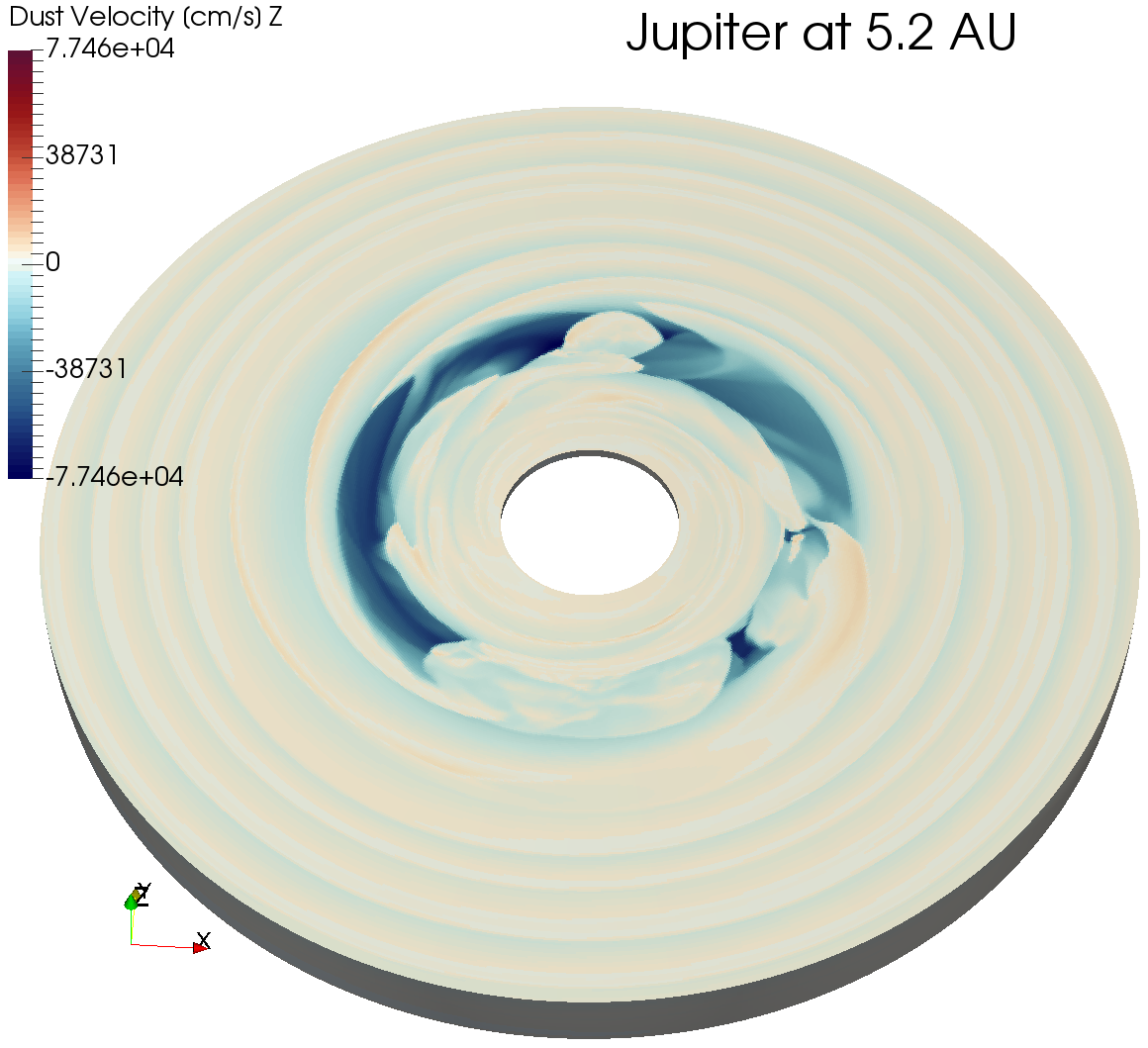}\includegraphics[width=9cm]{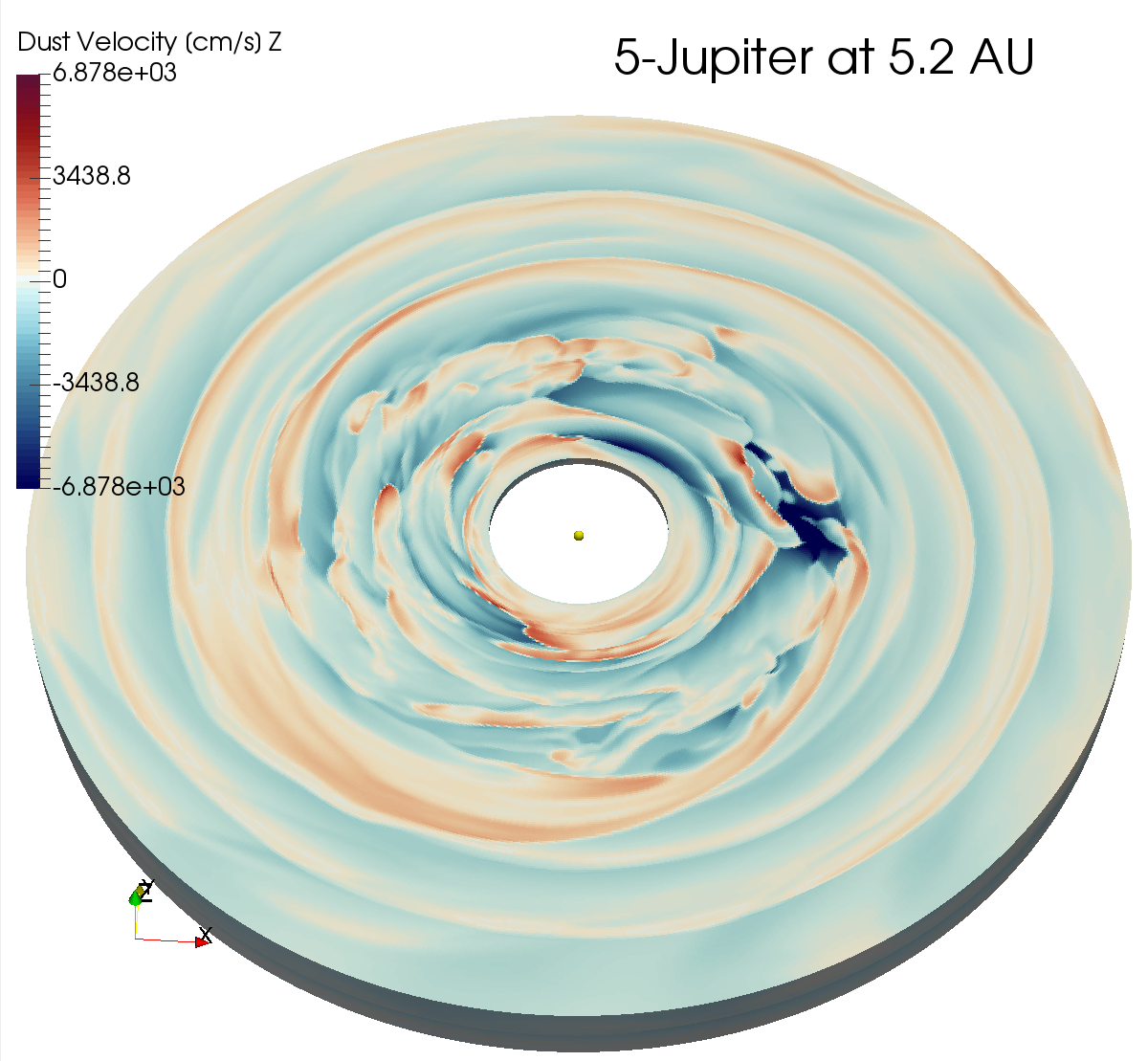}
\caption{Vertical velocities of dust in the circumstellar disk surface layer: the planetary spiral wakes transport material upward and downward inside the circumstellar disk. {The velocities reach higher absolute values with increasing planet mass, and we note that the overall hydrodynamic turbulence of the circumstellar disk also increases with planet mass.} There is no significant difference between the simulations with the planet at different semi-major axes. Hence, in this figure, we only show the simulations in which the planets are put at 5.2 AU. We only show one component of the three-dimensional velocity vector. Therefore, it is not representative of the actual flow. For the 3D streamlines, see Fig. \ref{fig:merid_circ_stream}.}
\label{fig:spirals}
\end{figure*}

\begin{figure*}
\includegraphics[width=18cm]{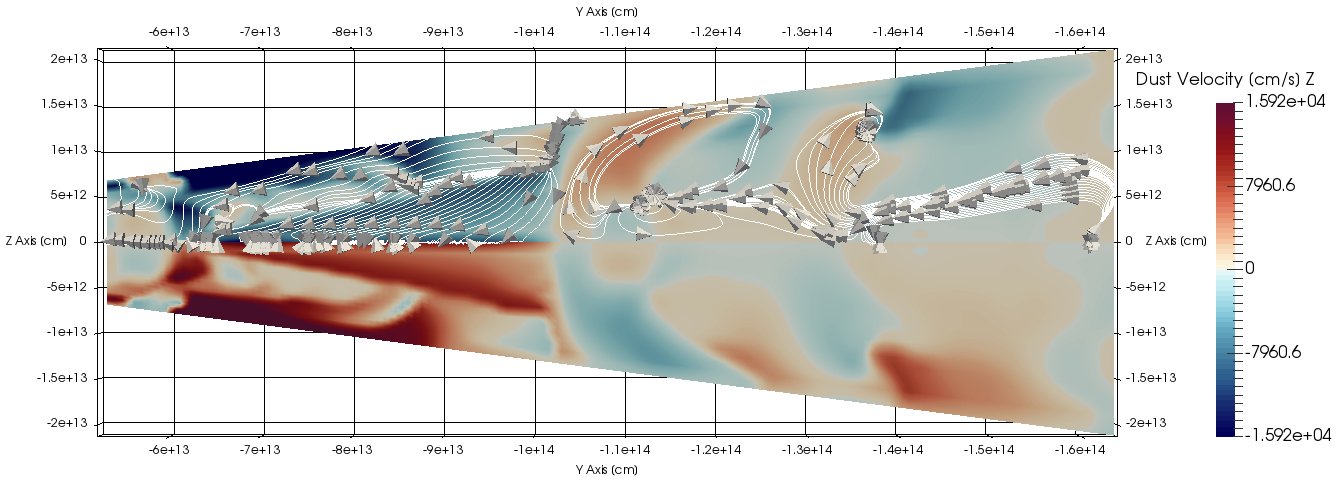}
\caption{Streamline plot projected onto a vertical slice 90 degrees away from the planet, which shows how the meridional circulation drives the dust flow: the spiral wakes (red-blue vertical stripes) circulate the dust from the midplane to higher altitude regions, and bring dust into the gap region from higher altitudes. The background colors represent the vertical component of the dust velocity. These data are from the 5 Jupiter-simulation at 50 AU.}
\label{fig:merid}
\end{figure*}

\begin{figure*}
\centering
\includegraphics[width=18cm]{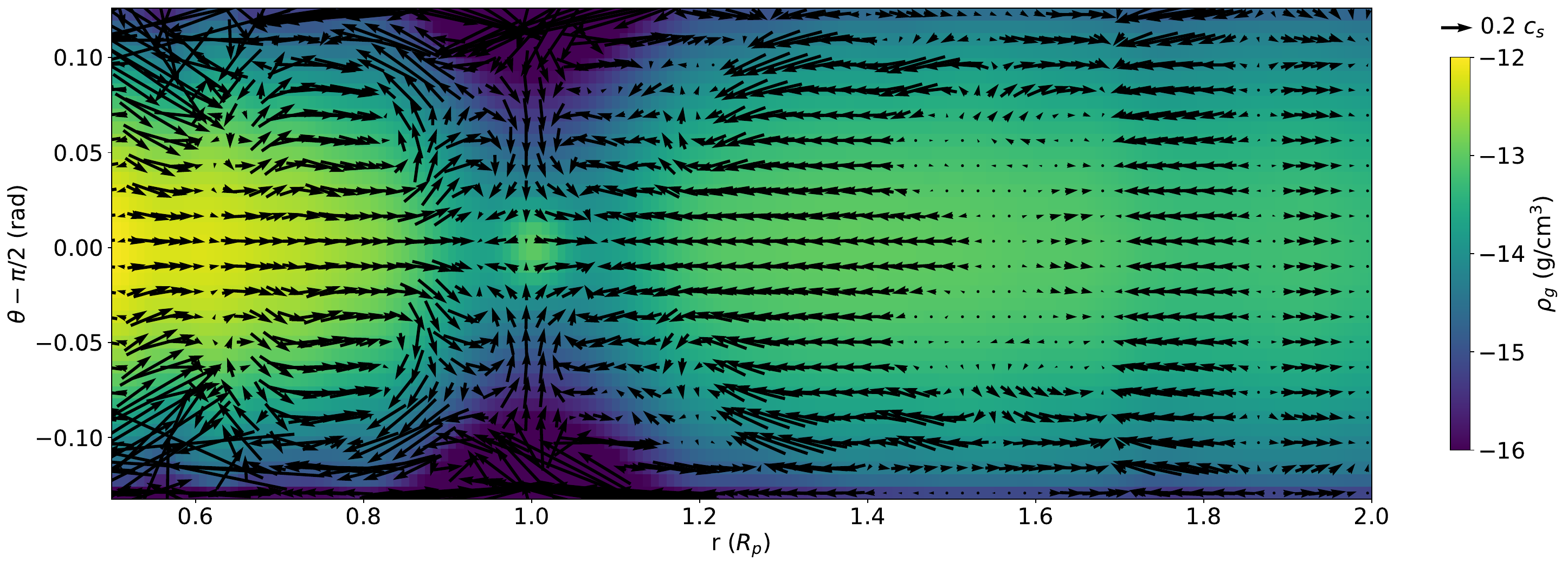}
\includegraphics[width=18cm]{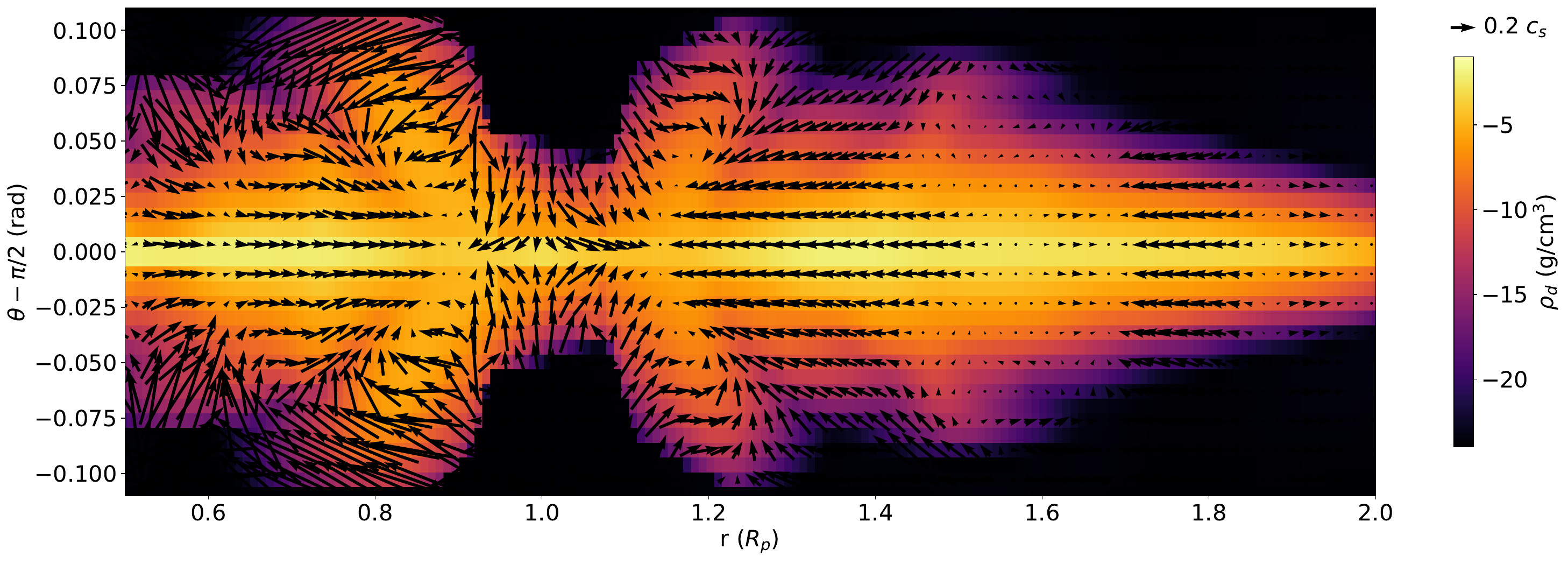}
\caption{{Azimuthally averaged (across $\phi=0\pm0.25$) velocity structure in gas (top) and in dust (bottom) of the radial and polar velocity components normalized to the local sound speed in the simulation containing a Jupiter-mass planet orbiting at 50 AU. For illustrative purposes, we present only every second velocity vector. The background contours show the gas volume density and the dust volume density respectively in a logarithmic color scale. Both velocity and density values represent a snapshot in time after 200 planetary orbits in the simulation. }}
\label{fig:quiver}
\end{figure*}

\begin{figure*}
\includegraphics[width=18cm]{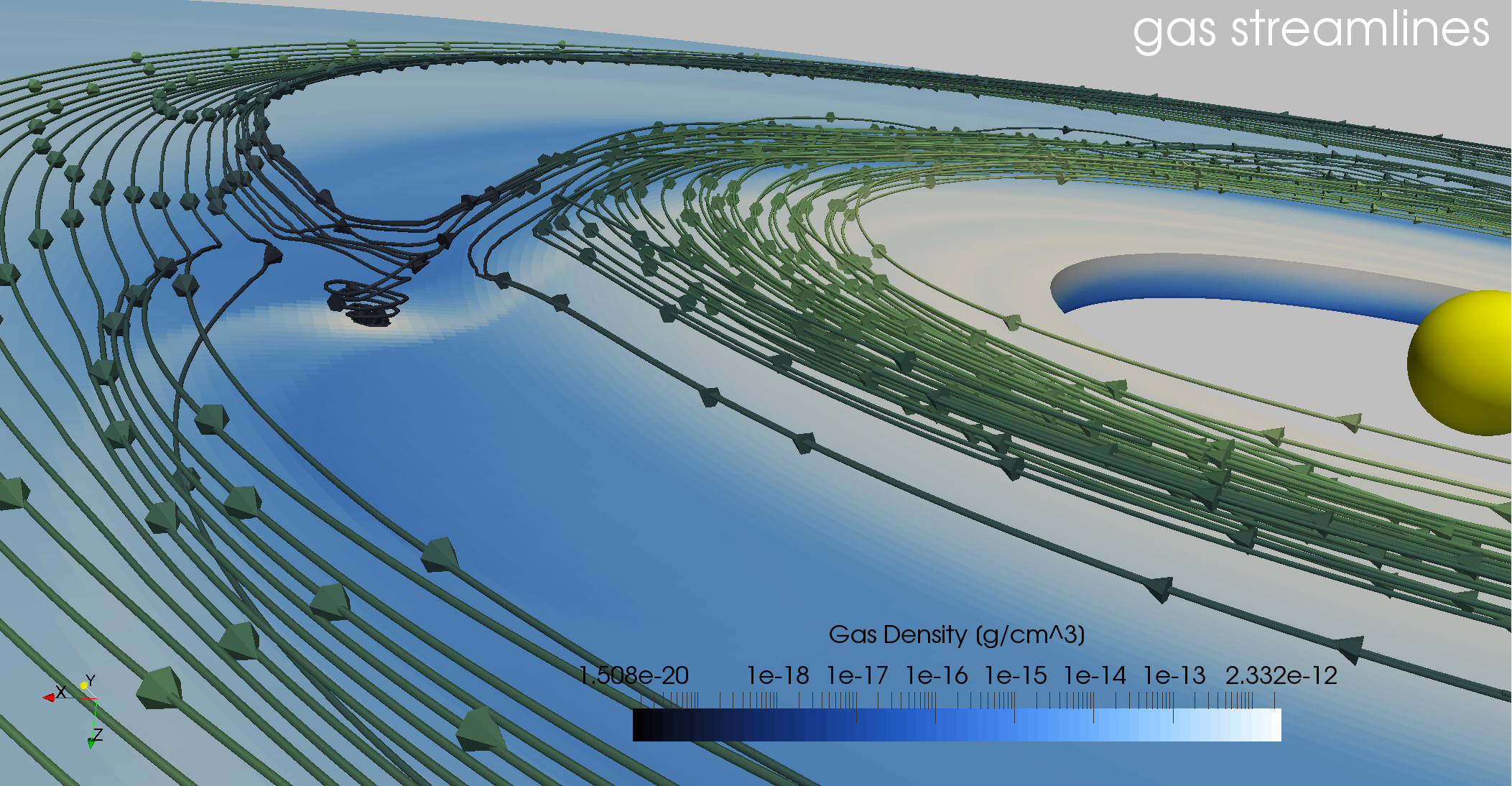}
\includegraphics[width=18cm]{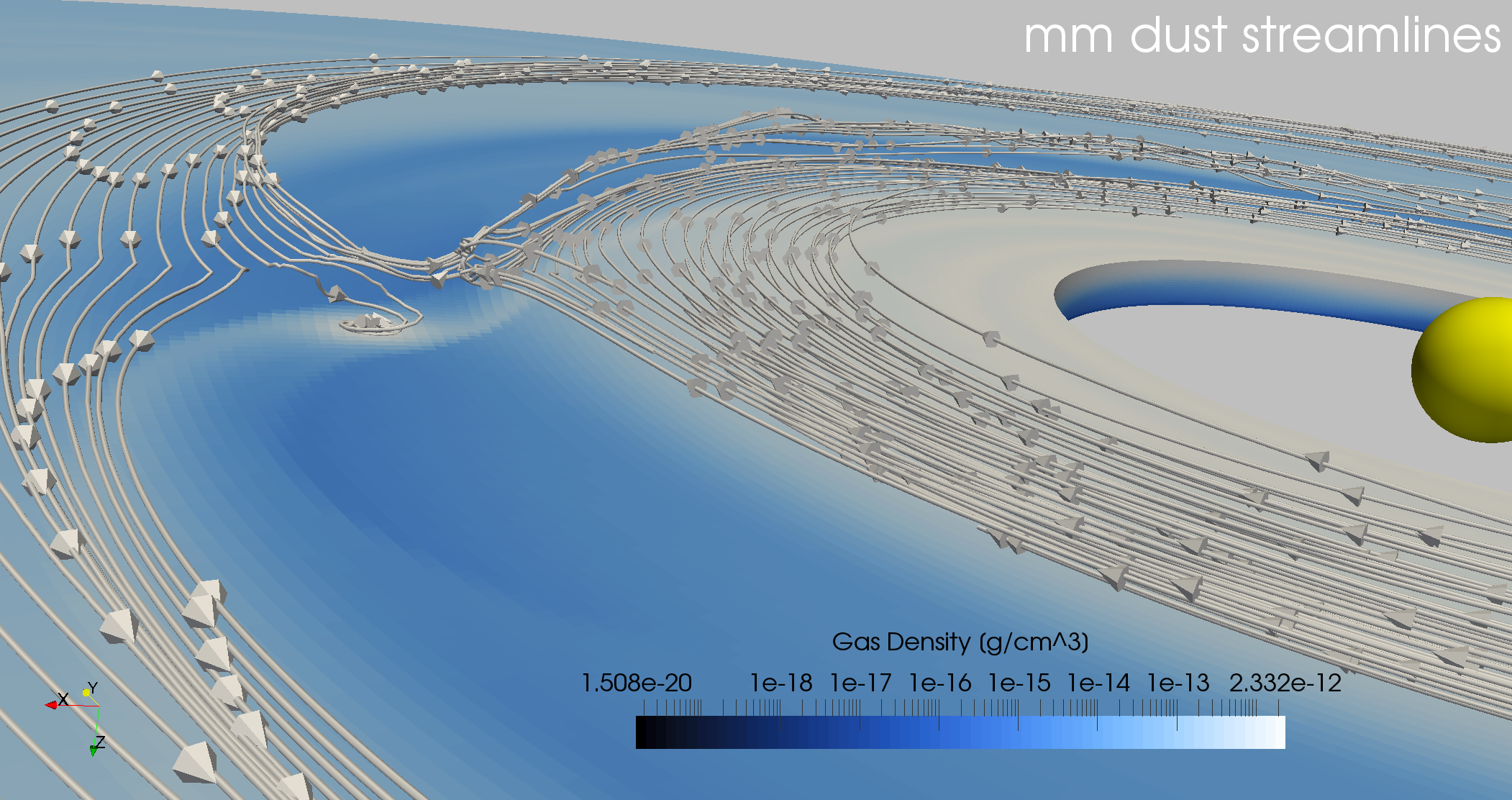}
\caption{Three-dimensional velocity streamlines for the gas flow (top) and for the mm-dust flow (bottom) in the case of the Jupiter-mass planet at 30 AU from the star. The mm-dust follows the gas flow well and the streamlines are similar. This means that the mm-dust grains mainly arrive in the circumplanetary region from the vertical direction, not through the midplane regions (see also Fig. \ref{fig:quiver}). This is due to the spiral shock waves, {that lift up the dust from the midplane and distribute} it at higher altitudes. The flow is very similar in simulations with different semi-major axes of the planet, hence we show only this one simulation. With increasing planetary mass, the meridional circulation is stronger, the spiral wakes create more substantial shocks that stir up the dust into higher regions of the circumstellar disk. The meridional circulation is clearly a very strongly three-dimensional flow, where the streamlines coming from the outer and inner circumstellar disk shock on the spiral wakes of the planet. Hence, they lose angular momentum and merge into a closer orbit toward the planet. This is how parts of the flow spirals down to the circumplanetary region from the higher circumstellar regions, while other parts turn back in the horseshoe region of the planetary gap.}
\label{fig:merid_circ_stream}
\end{figure*}

Here we examine how the spiral wakes of the planet stir up the dust and influence material flows. Fig. \ref{fig:spirals} shows the vertical velocities, {where it is strikingly visible that each spiral wake consists of a downward flow (blue) and an upward flow (red)}. We also show a representative streamline plot which shows a vertical slice of the dust velocities (Fig. \ref{fig:merid}), that visualizes how the planet generated spiral arms vertically mix the dust (and gas), and that the spiral wakes of planets are responsible for the meridional circulation, as it was found in \citet{Szulagyi14} and \citet{FC16}. {In Fig. \ref{fig:quiver} we additionally show a more zoomed-in view of the velocity structure in the vicinity of a Jupiter-mass planet in both gas and dust.} The velocities reach higher absolute values with increasing planetary mass, and we note that the overall hydrodynamic turbulence of the circumstellar disk also increases with the planet mass. Therefore, with increasing planet mass, the spiral wakes distribute the dust particles higher (further away from the midplane) in the circumstellar disk (Fig. \ref{fig:merid_circ_stream}), and in general, they reprocess larger fractions of the dust in the disk (the stirring happens farther away from the planet in the radial direction).

Fig. \ref{fig:spirals} shows only one (the vertical) velocity component of the 3D velocity vector, so it is not representative of the actual flow (i.e. it is not only locally upward and downward motion) at each spiral wake. The three-dimensional streamlines show (Fig. \ref{fig:merid_circ_stream}) the real motion of the gas, due to the meridional circulation. By looking at the streamlines, it becomes clear how the flow bridges over above the midplane, and deposits the material in the vicinity of the planet.

\begin{figure*}
\includegraphics[width=18cm]{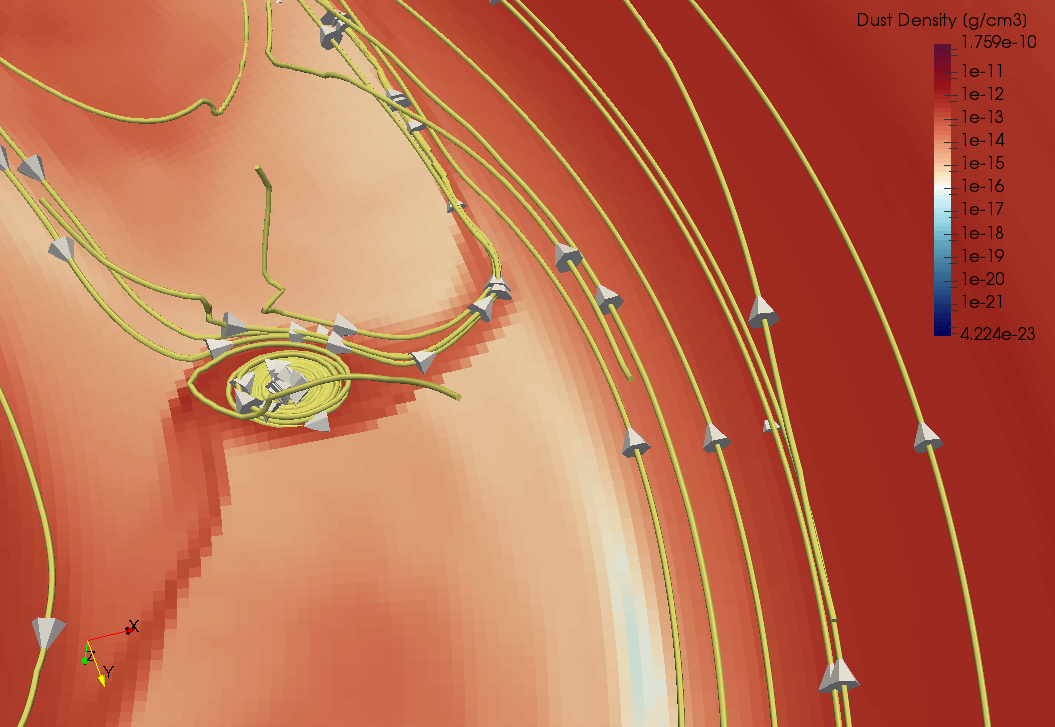}
\caption{Zoomed in view of the vicinity of the circumplanetary disk {showing} the three-dimensional dust velocity streamlines. Parts of the dusty-flow, which could not be accreted by the inner circumplanetary disk, is leaving this area through the spiral wakes and flows back to the circumstellar disk. This plot is from the 5 Jupiter-mass simulation at 5.2 AU.}
\label{fig:outflow_midplane}
\end{figure*}

\begin{figure*}
\centering
\includegraphics[width=18cm]{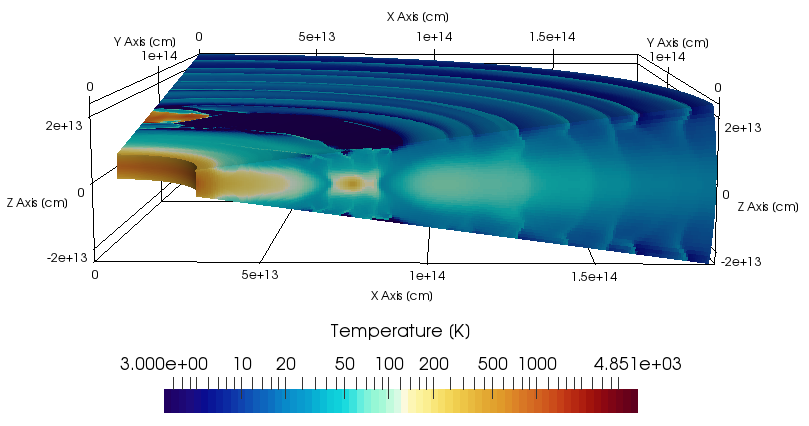}
\caption{Temperature field sliced at the planet location. This plot is from the 1 Jupiter-mass planet simulation placed at 5.2 AU from its star.}
\label{fig:temp}
\end{figure*}

\begin{figure*}
\centering
\includegraphics[height=6.5cm]{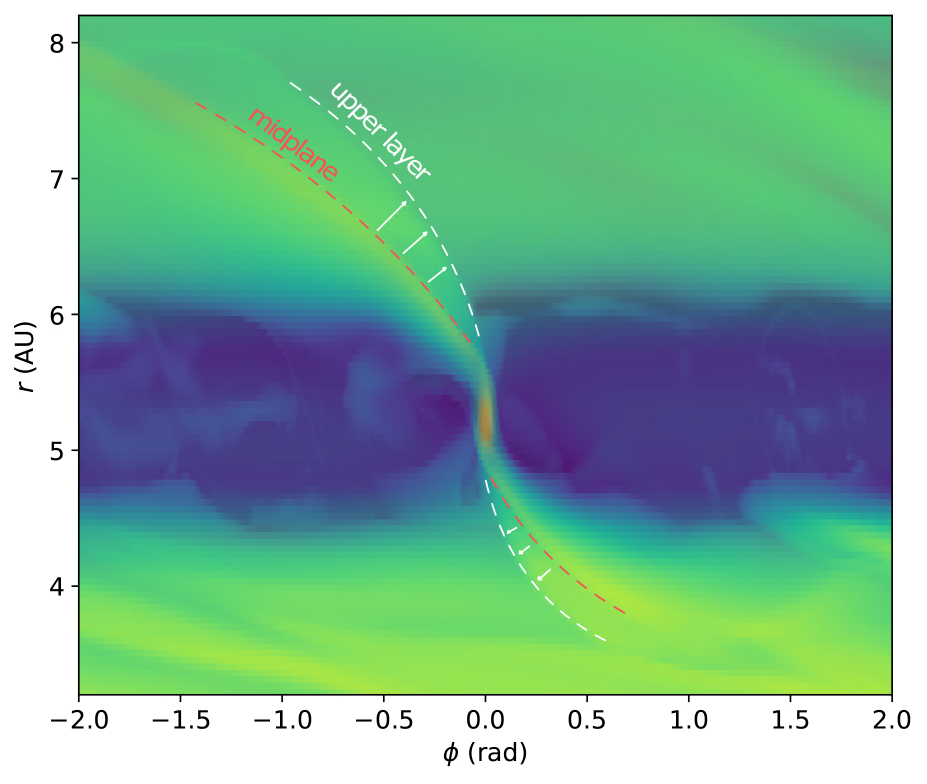}
\includegraphics[height=6.5cm]{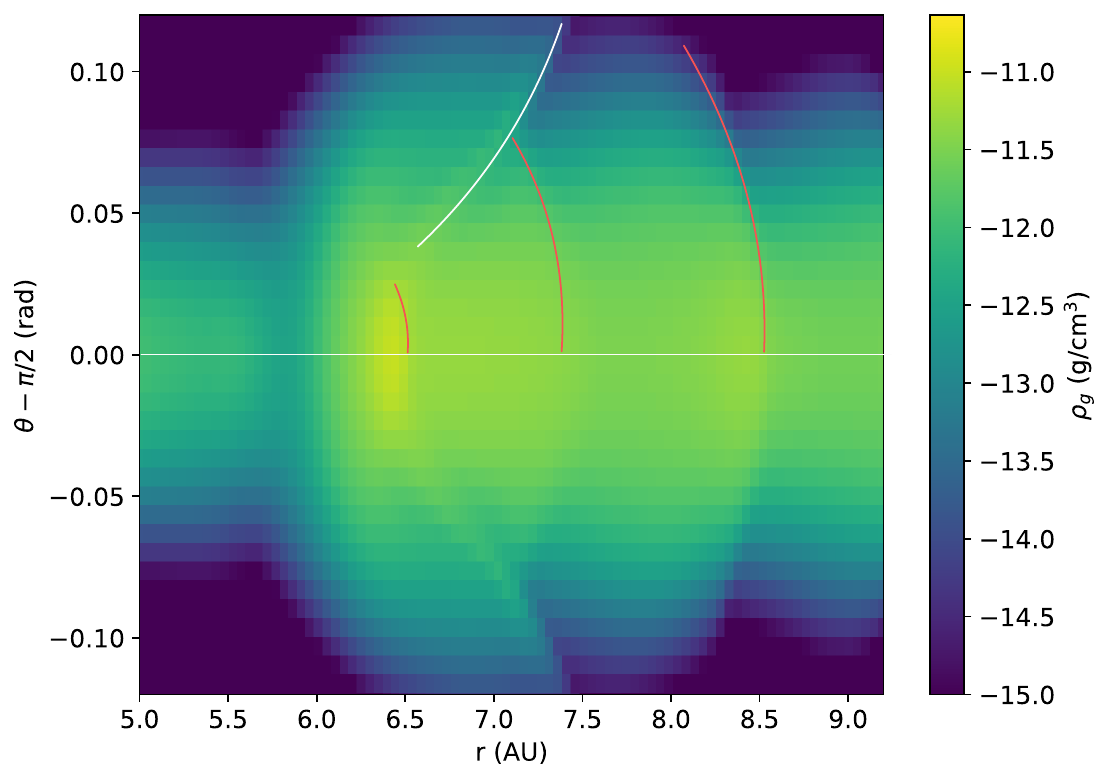}
\caption{In a realistic disk, the temperature, and consequently the sound speed, is not constant vertically. This leads to different pitch-angles of the planet wakes and different gap widths (similarly as it was discussed in \citet{Rosotti20}) and in \citet{Szulagyi17} ). The figures above show a Jupiter-mass planet at 5.2 AU and the gas density distribution in log-scale. Left: Plotted is an overlay of layer 18, i.e. a horizontal cut of an upper layer at 0.11 radians above the midplane, and the layer at the midplane. Here, we see that the pitch angle of the spiral arm is smaller in the midplane than in the upper layer of the circumstellar disk. This means that the pitch angle of the spiral wakes changes with increasing altitude from the midplane, which changes the dust's vertical distribution accordingly. {Right: Vertical cut in the $r-\theta$-plane at $\phi=-0.5$. Visible are several wave fronts at different radial distances from the planet. Indicated in red are Lindblad resonances, which have decreasing pitch angles with increasing distance to the midplane. Indicated in white is the wavefront of a buoyancy spiral. Notice that the midplane density in the left panel traces the Lindblad spiral, whereas the upper layer traces the buoyancy spiral. }}
\label{fig:pitchangel}
\end{figure*}

Our circumstellar disks, like most realistic disks, have a vertical temperature gradient (Fig. \ref{fig:temp}), therefore, we found the spiral wakes of the planet to have a different pitch angle at different distances from the midplane. 
This is especially pronounced when the planet is in the inner disk at 5.2 AU (Fig. \ref{fig:pitchangel}). We find the main spiral, close to the planet, to increase its pitch angle with increasing distance from the midplane. This is the opposite of what one expects from the spiral density waves induced by the planet, considering the generally increasing temperature with increasing distance to the midplane in our disks \citep[][]{Juhasz20, Rosotti20}. We find, close to the planet, spiral waves that are launched at buoyancy resonances \citep[][]{Lubow98,Zhu12b,Zhu15b} to interfere with the Lindblad spirals within a radial region $\pm1.5$ times the planetary orbital radius. {Since the vertical pressure gradient and the temperature gradient generally vanish as $z\rightarrow 0$, the midplane is void of buoyancy spirals and only the Lindblad spiral is visible (see midplane spiral in Fig. \ref{fig:pitchangel}). Above the midplane, buoyancy resonance appear and grow stronger with increasing distance to the midplane as previously shown in \cite{McNally20} for low mass planets. With increasing distance to the midplane, the most prominent buoyancy spiral, which has increasing pitch angle with increasing distance to the midplane, crosses the Lindblad spiral (which has decreasing pitch with decreasing distance to the midplane). Above this intersection, we find the buoyancy spiral to be more prominent than the Lindblad spiral. Overall, this results in a wavefront that has an increasing pitch angle with increasing distance from the midplane in the region where buoyancy waves are present.} Farther away from the planet, where the influence of buoyancy waves has dissipated, we find decreasing pitch angles with increasing distance from the midplane, consistent with our negative vertical temperature gradient. The overall effect is that, due to its vertical structure, the spiral wake opens like a fan, and this effect will also influence where the dust is distributed vertically in the disk. Moreover, it will possibly influence the observational appearance of the spiral arms. Firstly, because they will appear wider, and secondly, because the observed pitch angle will likely represent the opening angle of the spiral in the layer where there is optically thin to thick transition is, and not in the midplane. Based on our tests, in the not so realistic case, when the disk has no vertical temperature gradient (i.e. in locally isothermal hydrodynamic simulations), the spiral wake's pitch angle remains the same from the midplane to the surface of the disk.

As the planet orbits in the disk, with each passage, the spiral arms in the vicinity of the planet will stir up the local dust reservoir. When the planet has passed, the stirred-up dust starts to settle towards the midplane again. Each time, a new lobe settles onto the previous one, creating a layered structure. The dust settling timescale \citep{Birnstiel16} is inversely proportional to the Stokes-number ($St$) and the Keplerian angular velocity ($\Omega_k$): $\tau_{sett}=1/(St\,\Omega_k)$. Therefore, the settling will be more efficient in the outer disk.

\subsection{Dust and Gas Gaps of the Planet}
\label{sec:gaps}
\begin{figure*}
   \centering
\includegraphics[width=8.5cm]{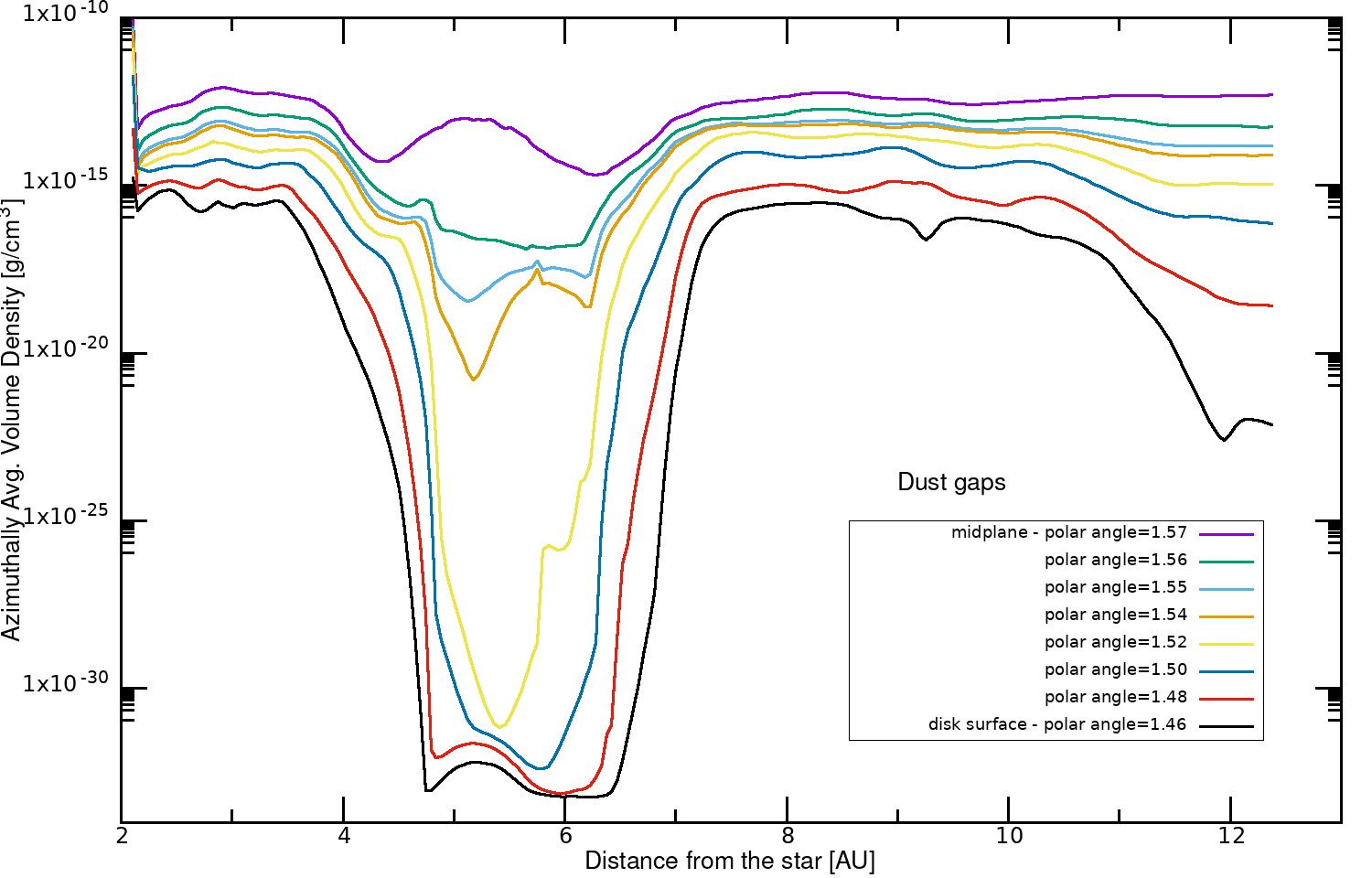} 
\includegraphics[width=8.5cm]{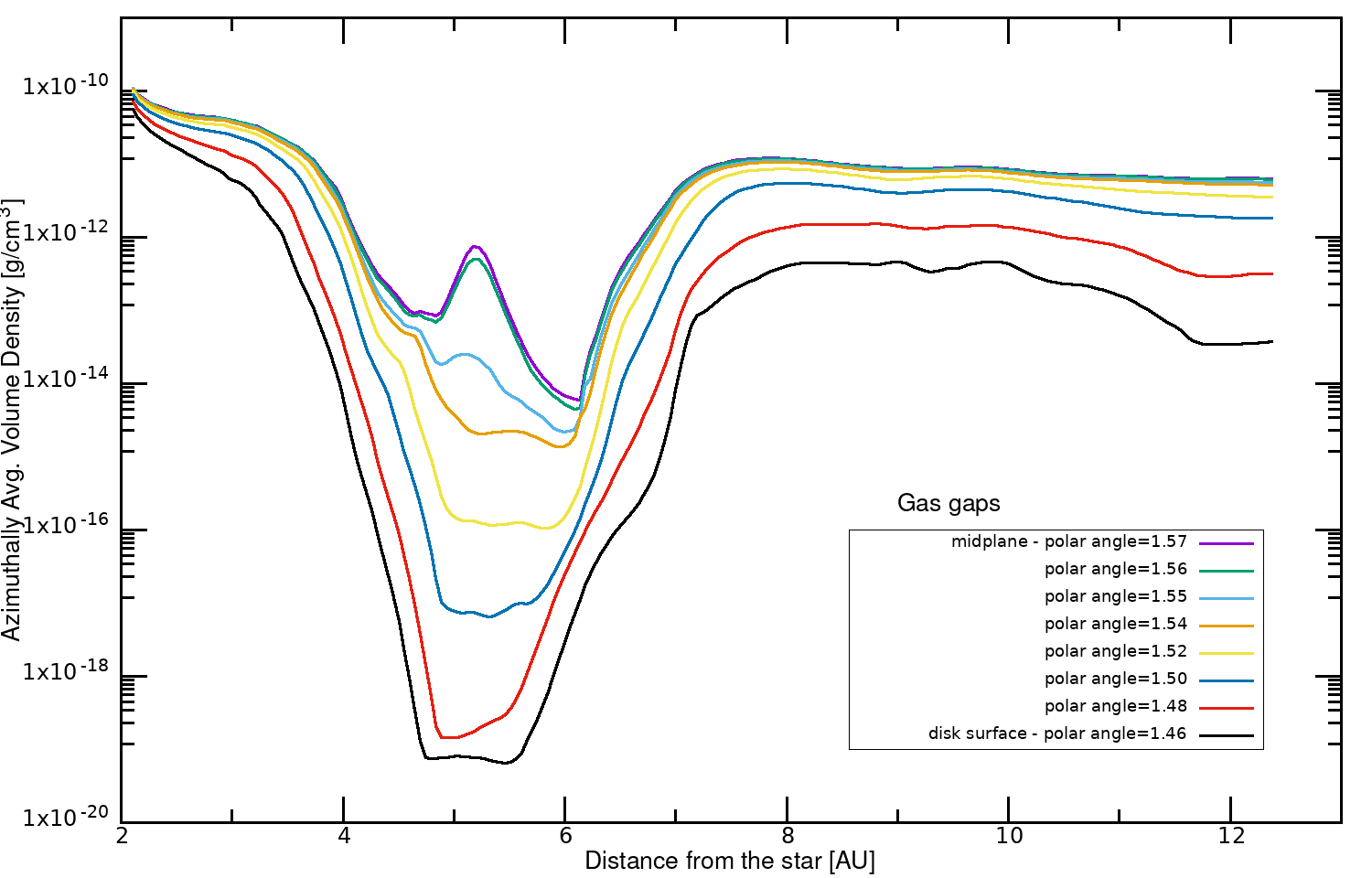} 
\caption{Gap-profiles for dust (left) and gas (gas) at different vertical distances to the midplane. Due to the vertical temperature gradient of the disk, the gaps are wider in the upper disk than at the midplane (see also in \citet{Szulagyi17}). These azimuthally averaged density profiles are from the 5 Jupiter simulation at 5.2 AU.}
\label{fig:gaps}
\end{figure*}

Vertically, the planetary gap does not have the same width (Fig. \ref{fig:gaps}). This is due to the vertical temperature gradient and was discussed for gas gaps in \citet{Szulagyi17}. Vertically isothermal hydrodynamic simulations, that per definition do not have a vertical temperature gradient, will result in a constant gap width from the midplane to the top of the circumstellar disk \citep{Masset16}. However, any realistic disk has a vertical temperature gradient, hence temperature included, radiative hydrodynamics are needed to understand the vertical structure of gaps and rings, which will affect the observations.

\subsection{Dust Stirring and the Delivery of Solids to the Circumplanetary Region}
\label{sec:stirring}

The meridional circulation \citep{Szulagyi14,FC16} stirs up the dust in the circumstellar disk (driven by the spiral wakes of the planet ) as shown in Figs. \ref{fig:spirals}, \ref{fig:merid_circ_stream}, \ref{fig:outflow_midplane}. Planets always trigger spiral wakes in a gaseous circumstellar disk, hence this meridional circulation mechanism is robust and the vertical stirring of the dust always happens when a planet is present. We found that the larger the planetary mass, the stronger is the stirring, the puffier is the dust disk and the higher is the altitude to which dust is delivered. This trend is easy to understand since the larger the planet mass is, the stronger spiral wakes are and the larger the vertical velocity within them is. Hence, the vertical stirring of the dust is more significant (higher, farther from the planet in radial direction, and a larger amount of dust is recycled). Here we simulated mm-sized grains, that were previously thought to be quite well settled to the midplane. Our finding is that when forming planets are more massive than Neptune (closer to Saturn-mass), the vertical stirring for mm-sized grains is quite strong, and maybe detectable by observations. The stirring is continuous once the planet is massive enough and the dust does not have time to settle back to the midplane. We found the vertical extent of the dust disk to be almost constant (Fig. \ref{fig:dustvertical}). As we described in Sect. \ref{sec:spiral}, with each passage of the planet, additional dust is stirred up, and therefore, new dust layers form.

We ran simulations with various semi-major axes of the planets (5.2 AU, 30 AU, 50 AU). Interestingly, we found that there are only small difference between simulations with different semi-major axes in terms of this spiral arm driven dust stirring. Given that the vertical extent of the circumstellar disk is smaller closer to the star, the planets that lie closer in (e.g. at 5AU vs. 30 AU) will be able to bring the dust more efficiently onto the surface of the circumstellar disk than planets further away, where the disk vertical extent is larger. Naturally, there is some dependence on Stokes number, and hence coupling as well, which is generally smaller (more well coupled) in the inner circumstellar disk. Finally, a planet closer to the star will experience stronger torques \citep{Rafikov02a,Rafikov02b}, hence the planet's spiral wakes will be somewhat stronger, than an equivalent planet with larger orbital separation. We checked the vertical velocity peak values in all simulations, and indeed, the peak value is higher for planets closer to the star. Albeit all these, we did find similar amount of stirring at the different orbital distances for the same mass planet, although there is a weak trend with orbital separation: {the closer the planet is to the star, the more stirring that occurs (see also \citealt{Binkert21}).}

We derived the net mass gain of the Hill-sphere (i.e. calculating the difference between inflow and outflow). The circumplanetary disk is a substructure of the Hill-sphere, but our simulations do not resolve the CPD well, hence we do not determine mass fluxes well within the unresolved Hill-sphere because those numbers would not be correct. We measured the net mass gain over the last orbit of the planet and converted it to physical units ($\rm{M_{Jup}/yr}$). The net mass gain of the Hill-sphere, broken down by species (mm-dust and gas), is reported in Table \ref{tab:influx}. The values are always positive, which means the influx is higher than the outflow, hence the Hill-sphere seems to become enriched with mm-sized grains. The generic trend, as seen in Table \ref{tab:influx}, is that the higher the planetary mass is, the higher the dust- and gas-accretion rates are. The net dust-mass gain of the Hill-sphere is on the order of $10^{-8}$ to $10^{-10}$ $\rm{M_{Jup}/yr}$. For the gas, it is $10^{-6}$ to $10^{-8}$ $\rm{M_{Jup}/yr}$. There are usually one to two orders of magnitude difference between the two species. This mass influx difference between the dust and gas is smaller for smaller mass planets. There is no obvious trend for the accretion rates of the Hill-sphere when with different semi-major axes. This means, that the CPD is not dust starved \citep{Zhu18}, but dust rich \citep{DSz18}. The CPD is constantly fed by the circumstellar disk not only with gas, but also with micron-sized and mm-sized particles, which allows moon-formation to eventually occur. This also shows that the pressure maximum outside the gap does not prevent dust grains from reaching the circumplanetary disk (or envelope) since they are delivered by the stirring of the spiral wakes from above the midplane. An additional consequence is that there is no need for planetesimal capture from the circumstellar disk to form satellites \citep{Fujita13,Tanigawa14,DAngelo15,Suetsugu16,Ronnet20}, but dust coagulation can occur in the CPD \citep{DSz18}, opening the way for moon-formation \citep{Cilibrasi18,Szulagyi18b,Shibaike19,Inderbitzi20,Batygin20,Cilibrasi21}. Comparing our accretion rates with literature values, for mm-sized dust, \citet{Homma20} found between $5\times10^{-9}$ and $5\times10^{-6}$  $\rm{M_{Jup}/yr}$ in the case of a Jupiter analogue, which is a fitting range to our $1.6\times10^{-8}$ value. Regarding the gas accretion rate onto the Hill-sphere, in our equivalent isothermal simulation for a Jupiter analogue in \citep{Szulagyi14}, we found a few times $\times10^{-6}$  $\rm{M_{Jup}/yr}$, which is an order of magnitude larger value than in our current radiative simulations. \citet{AB09} found in their radiative simulations, planetary accretion rates (i.e. not Hill-sphere influx rate) which are on the order of few times $\times10^{-5}$ $\rm{M_{Jup}/yr}$, depending on the assumed opacity. Similarly, \citet{Tanigawa16} calculated a planetary gas accretion rate from isothermal simulations for a Jupiter-mass planet, which is a few times $\times10^{-5}$  $\rm{M_{Jup}/yr}$. Differences in the circumstellar disk density profile, and in the accretion rate calculations, in opacities, and in temperature treatment result in these differences.

\begin{table*}
\begin{center}
\begin{tabular}{l| c |c |c |c} 
 \hline
   simulation & $\rm{M_{p}}$ [$\rm{M_{Jup}}$] & d [AU] &  ${{\dot{M}_{Hill_{dust}}}}$ [$\rm{M_{Jup}/yr}$]  &   ${{\dot{M}_{Hill_{gas}}}}$ [$\rm{M_{Jup}/yr}$]  \\ 
  \hline
m5au1nep & 0.1 &  5.2 &  1.1e-09 &   9.1e-09  \\
m5au1sat  & 0.3 &  5.2 &  7.9e-09 &    1.4e-07 \\
m5au1jup &  1.0 &  5.2 &  1.6e-08 &   5.0e-07  \\
m5au5jup &  5.0 &  5.2 &  3.5e-09 &   2.5e-06  \\
m30au1nep &  0.1 & 30.0 &  7.0e-10 &   2.0e-09 \\
m30au1sat &  0.3 & 30.0 &  6.4e-10 &    8.5e-08 \\
m30au1jup &  1.0 & 30.0 &  3.5e-10 &    4.3e-07  \\
m30au5jup &  5.0 & 30.0 &  7.3e-08 &   1.8e-06  \\
m50au1nep &  0.1 & 50.0 &  4.8e-10 &   9.8e-10  \\
m50au1sat &  0.3 & 50.0 &  3.1e-10 &  5.8e-08 \\
m50au1jup &   1.0 & 50.0 &  9.7e-08 &    7.2e-06  \\
m50au5jup &   5.0 & 50.0 &  6.5e-08 &    3.5e-07 \\
 \hline
\end{tabular}
\caption{Delivery onto the circumplanetary region. Net mass gain of the Hill-sphere for the mm-sized dust and for the gas.}
\label{tab:influx}
\end{center}
\end{table*}

\subsection{Local Dust-to-Gas Ratio}
\label{sec:dtg}

\begin{figure*}
\begin{tabular}{|c|}
\hline
\includegraphics[width=18cm]{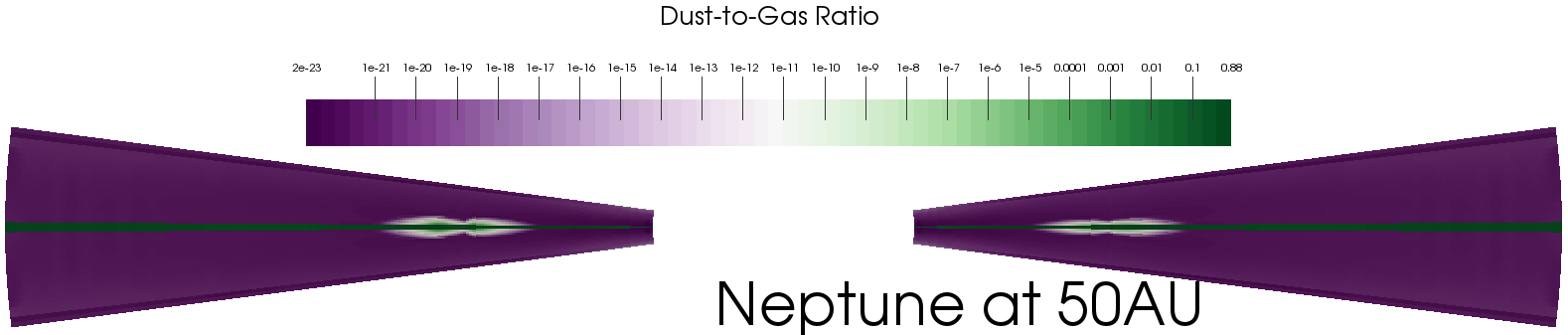} \\
\hline
\includegraphics[width=18cm]{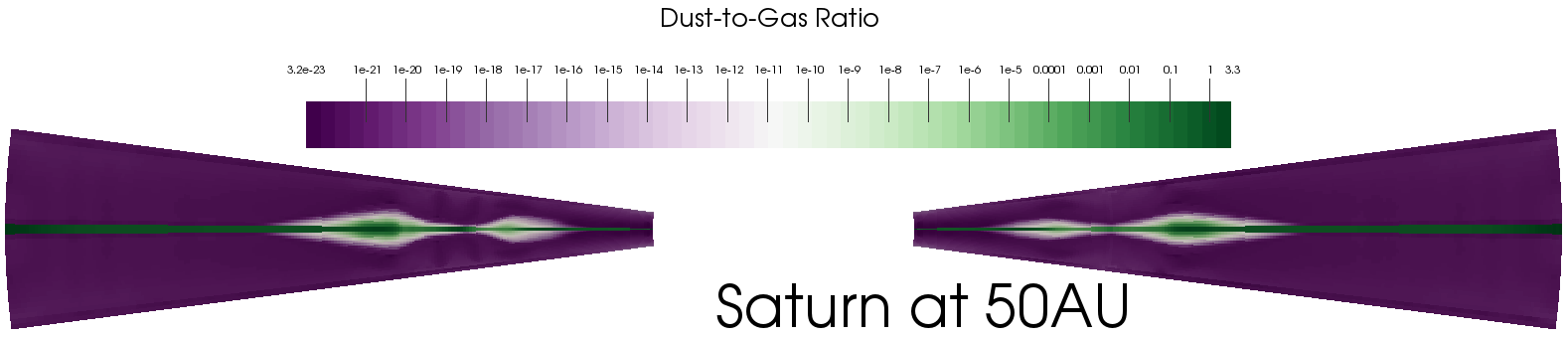}  \\
\hline
\includegraphics[width=18cm]{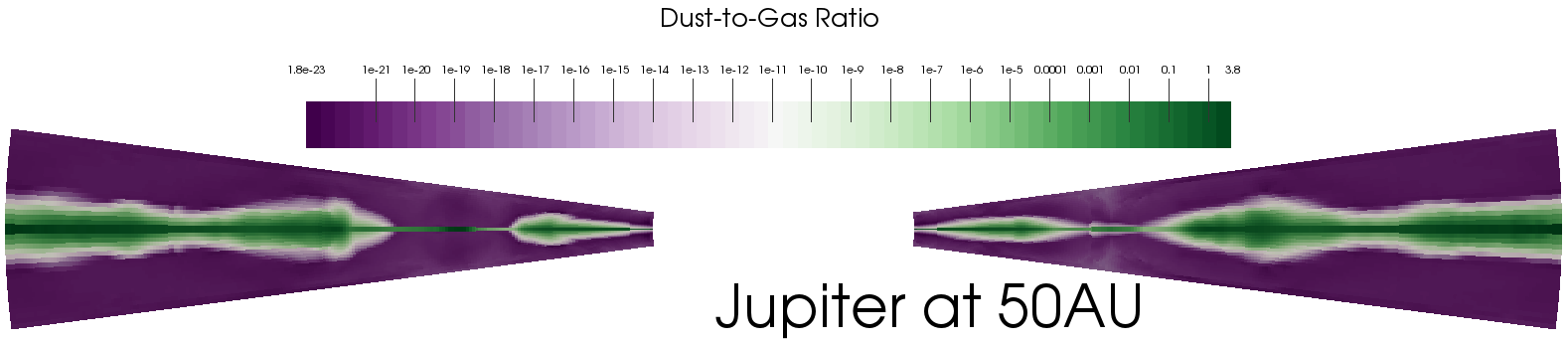} \\
\hline
\includegraphics[width=18cm]{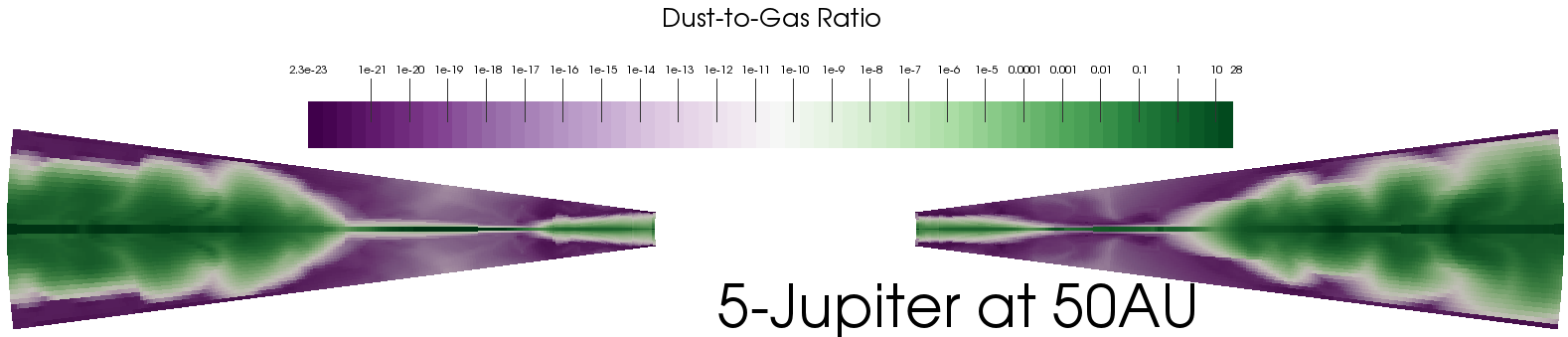}  \\
\hline
\end{tabular}
\caption{Vertical slices of the local dust-to-gas ratio at the location of the planet and opposite to the planet. There is no significant azimuthal variation in the local dust-to-gas ratio. The panels show the Neptune-, Saturn-, Jupiter-, and 5 Jupiter-mass planets (from top to bottom) that are orbiting 50AU from their star. Clearly, the dust stirring is stronger with larger planetary mass (regardless of the planet's semi-major axis). The larger the planetary mass is, the higher the mm-dust is stirred up above the midplane. While in the simulation containing the Neptune-mass planet the dust is mainly confined to the midplane, the dust is heavily stirred up by the spiral wakes of the planet in the simulations containing a Saturn-mass planet and above.}
\label{fig:dust-to-gas}
\end{figure*}

\begin{figure*}
\includegraphics[width=9cm]{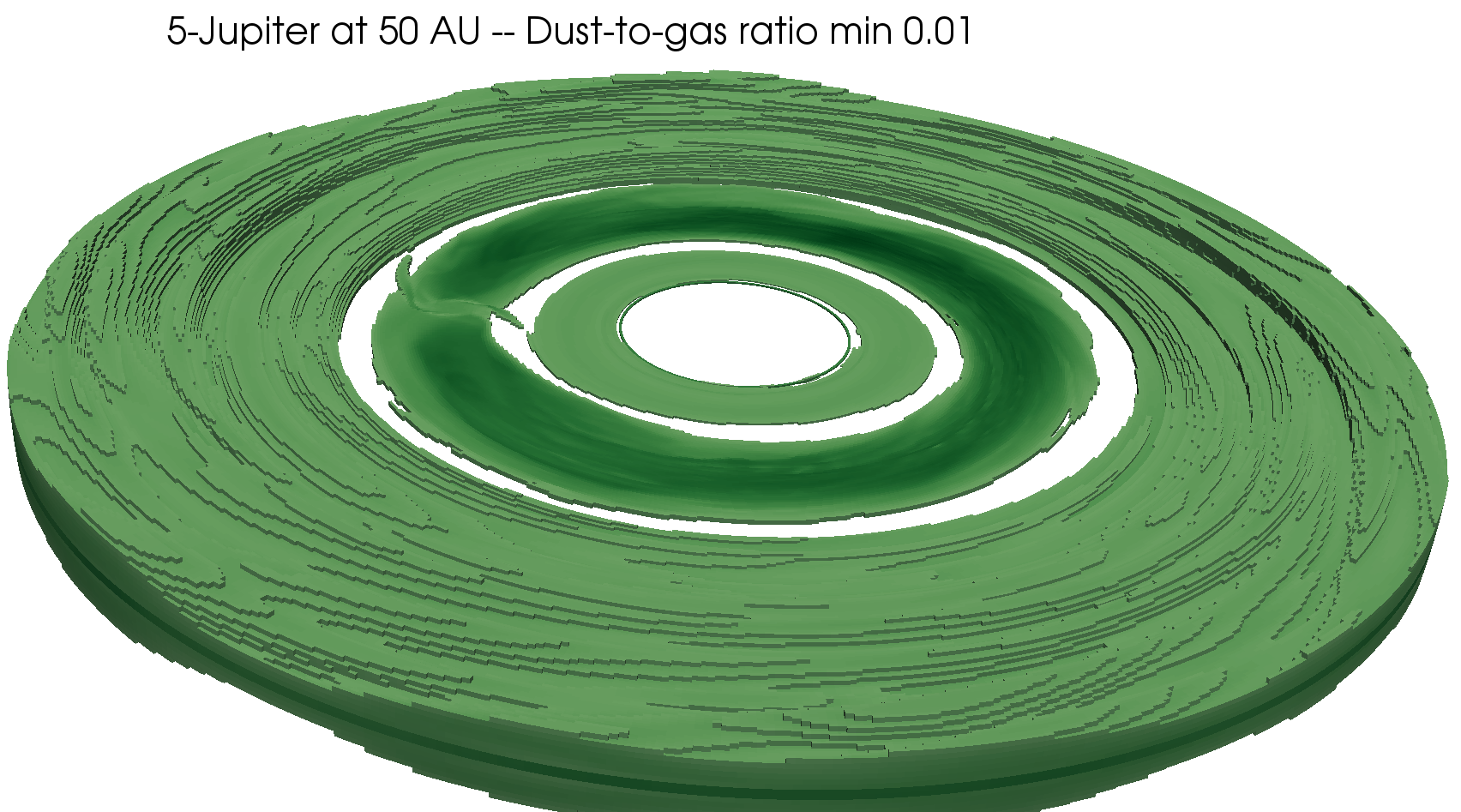}\includegraphics[width=9cm]{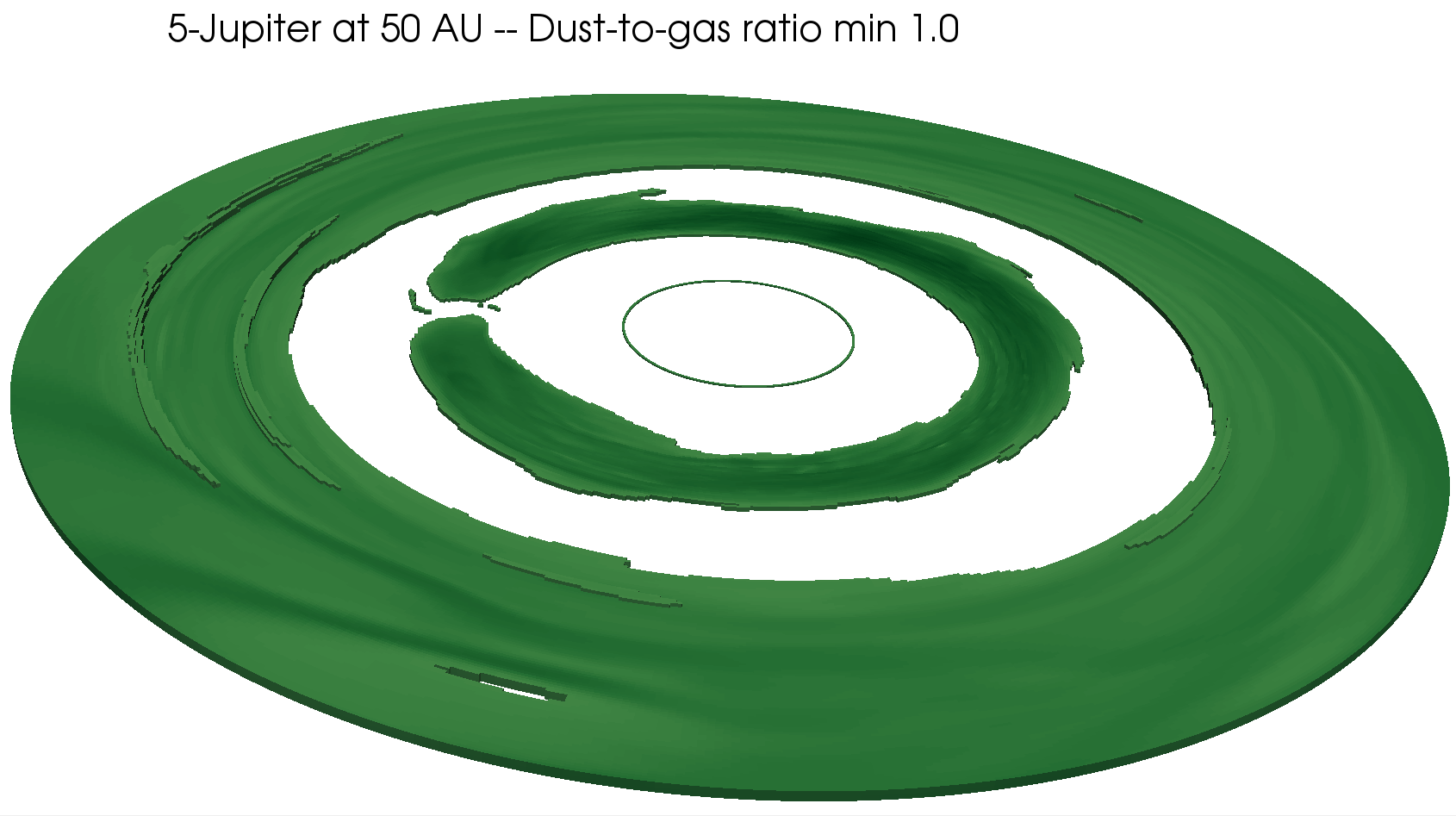} 
\caption{Dust-to-gas ratio surfaces shown with a threshold of 0.01 (left) and 1.0 (right) in the case of the 5-Jupiter-mass planet at 50 AU. The {DTG=1.0}-surface is still mainly confined to the midplane, however, the planet's spiral wakes accumulate dust as seen on the right panel. The {DTG=0.01}-surface is well above the midplane, mainly in the outer circumstellar disk (beyond the planet's orbit). As the lower-mass planets stir up the mm-grains less efficiently, {DTG=0.01}-surface lies closer to the midplane in those cases.}
\label{fig:dust-to-gas5jup}
\end{figure*}

We calculated the local dust-to-gas ratio in the entire circumstellar disk by dividing the dust- and gas volume densities in every cell (Fig. \ref{fig:dust-to-gas}). From the figure, it becomes apparent that the local dust-to-gas ratio varies considerably throughout the disk. It varies from gas-only regions where the local dust-to-gas ratio is close to $10^{-23}$ to values of up to $\sim$27 in other regions. The spiral wakes are always dust-rich, and they enhance the local dust density (Fig. \ref{fig:dust-to-gas5jup}). The maximum values of dust-to-gas ratios can be found in the midplane.

Fig. \ref{fig:dust-to-gas5jup} shows the 0.01 (left) and the 1.0 (right) dust-to-gas density ratio surface in the case of the 5-Jupiter-mass planet at 50 AU from the star. Even though the {1.0} ratio is still mainly confined to the midplane, the planet's spiral wakes are accumulating dust, as seen on the right panel. The {0.01} dust-to-gas ratio surface is well above the midplane, mainly in the outer circumstellar disk (beyond the planet's orbit). For smaller mass planets, the {1.0} ratio for mm-size grains is in the midplane, since they cannot stir up that much dust as the higher mass planets can.  

We determined the largest distance above the midplane (expressed in degrees), where the dust-to-gas ratio (DTG) is above 1.0, 0.1, 0.01 and 0.001 respectively for the different planets and semi-major axes (Table \ref{tab:dtg}). When there is no value given in the table (mainly for Neptune and sometimes Saturn-mass planets), it means that the DTG is always smaller in the entire circumstellar disk than the threshold value. For example, we never found DTG=1 or higher when the planet is only Neptune-mass. This again shows that there is larger local dust accumulation when the planet masses are higher: the spiral wakes are stronger and the pressure bumps are deeper, both of which help to enhance the local dust density. From Table \ref{tab:dtg}, it is clear again that the larger mass planets stir mm-dust to larger distances above the midplane. For Jupiter-mass planets, and more massive giants, the streaming instability condition of 1.0 dust-to-gas ratio is found also above the midplane. While the dust-rich surfaces are, at most, a few degrees above the midplane. Overall, we found in the previous paper, \citet{Binkert21} that this has a significant effect on e.g. the observations with ALMA. A large fraction of dust is hidden inside the optically thick regions of the disk, causing disk mass estimates based on ALMA continuum observations to significantly underestimate disk masses (see details in \citealt{Binkert21}). 

From Table \ref{tab:dtg} it is clear that there is no significant difference in the DTG and the surface thresholds between the different simulations with planets at different semi-major axis. How high the dust is distributed mainly depends on the planetary mass, not the planetary orbital distance. Hence, the vertical mixing of dust into regions above the midplane is robust and independent of the orbital separation between the planet and the star.

\begin{table*}
\begin{center}
\begin{tabular}{l c c c c} 
 \hline
   simulation & surf. DTG=1 [deg] &  surf. DTG=0.1 [deg]&  surf. DTG=0.01 [deg]  &  surf.DTG=0.001 [deg]  \\ 
  \hline
m5au1nep &  - &  0.19 &  0.19 &  0.56   \\
m5au1sat &  - &  0.19 &  0.56 &  1.67   \\
m5au1jup &  1.67 &  3.89 &  5.00 &  6.85   \\
m5au5jup &  0.19 &  3.15 &  4.63 &  6.85  \\
m30au1nep &  - &  0.19 &  0.19 &  0.56 \\
m30au1sat &  0.19 &  0.19 &  0.56 &  0.56  \\
m30au1jup &  0.19 &  0.19 &  1.67 &  2.41  \\
m30au5jup &  0.93 &  2.78 &  3.89 &  5.00 \\
m50au1nep &  - &  0.19 &  0.56 &  0.56 \\
m50au1sat &  0.19 &  0.19 &  0.56 &  0.56  \\
m50au1jup &  0.19 &  2.41 &  2.78 &  3.15  \\
m50au5jup &  1.30 &  3.52 &  4.26 &  5.00  \\
 \hline
\end{tabular}
\caption{The largest distance above the midplane (expressed in degrees) where the dust-to-gas ratio (DTG) is at least 1.0, 0.1, 0.01, 0.001, respectively.}
\label{tab:dtg}
\end{center}
\end{table*}

\section{Discussion}

The vertical dust stirring due to the presence of a massive planet changes the picture of our understanding of dust processes in circumstellar disks. Without a planet, and in the absence of other sources of turbulence, dust does settle onto the disk midplane. With a planet, dust is stirred up considerably above the midplane. This stirring is stronger, i.e. the vertical dust distribution is wider, and a larger radial fraction of the disk dust is stirred, the larger the planet mass is. This likely affects the observations of dust in the disk \citep{Birnstiel18, Dullemond18,Zhang18,Sz18} and the degree of dust stirring might give a clue about the presence, or absence, of a planet. Furthermore, dust stirring will influence disk chemistry \citep{Visser09,Visser11,Bergner20} and likely even chondrule formation \citep{Ciesla02,Alexander08,Dullemond14,Bodenan20}. Therefore, further investigation is necessary in the field to understand how the dust stirring affects all these processes. 

Our models are however limited. We only included one dust grain size, due to the extensive running-time of these simulations. When studying streaming instability, \citet{Krapp19} found that the process is significantly altered if multiple dust grain sizes together are simulated. It will be a necessary future study to investigate how the meridional circulation changes if multiple dust species are included in the simulations. Our simulations also lack dust turbulent diffusion. This does affect the dust distribution \citep{Zhu12,deJuanOvelar16,Krijt16}. For example, it increases the vertical stirring of the dust on top of the vertical stirring by the planet and counteracts the vertical settling of dust. We briefly tested the effect of dust turbulent diffusion in \citet{Binkert21}, and we will specifically study its effects on the vertical dust distribution and the dust delivery to the Hill-sphere in our upcoming paper. We also did not include dust coagulation and fragmentation, which certainly has a large effect on the meridional circulation of dust. Today, it is not yet possible to run such computationally expensive simulations which capture the full extent of dust growth in three dimensions. To our knowledge, only one group has done a 2D study \citep{Drazkowska19}. They found that the inclusion dust growth has an effect, e.g. on the gap width and pressure bumps. In the future it should be investigated, how dust growth influences the three-dimensional dust distribution circumstellar disks.

We did not add magnetic fields in these simulations. Magnetic fields can further influence dust stirring due to higher levels of turbulence e.g. due to magneto-rotational instability, and magnetic fields also affect the disk general flows. Magnetic fields can also drive winds that affect angular momentum transport which in turn affects mass transport in the disk.

In \citet{Binkert21} we also discussed that there are differences between isothermal and radiative hydrodynamic simulations for the vertical stirring of the dust, which shows that thermodynamic effects matter for this process too. We stress that we must resort to radiative hydrodynamics to realistically study these processes.

\section{Conclusions}

We carried out the first sets of three-dimensional dust+gas radiative hydrodynamic simulations of embedded planets with a grid-based code. Twelve different setups were examined: a Neptune-, a Saturn-, a Jupiter-, and a 5 Jupiter-mass planet orbiting at 5.2 AU, 30 AU and 50 AU from their Solar-equivalent star respectively. We included all relevant heating/cooling mechanisms in the simulations (stellar irradiation, viscous heating, accretion heating of the gas, radiative dissipation, adiabatic expansion and compression, shock-heating). The initial global dust-to-gas ratio was 1\%, and the dust composition in the calculation of the opacity and internal density was assumed to be 40\% silicate, 40\% water, 20\% carbon, with the sublimation of the species taken into account and gas opacities also included. We simulated mm-sized grains with a multi-fluid approach where both the gas and the dust dynamics have feedback on each other. 

Our findings can be summarized as follows:
\begin{itemize}
    \item Due to the presence of a planet, dust is stirred up vertically well above the midplane. This stirring is due to the meridional circulation \citep{Szulagyi14,FC16}, where the spiral wakes of the planet bring up the material from the midplane. 
    \item The stirring becomes stronger with increasing planetary mass: a larger fraction of dust is lifted, it is distributed higher in the disk, and the effect plays a role in a a farther radial distance from the planet.  The stirring is negligible for Neptune-mass planets and below (for the considered mm-sized grains), but important for Saturn-mass planets and above.
    \item The meridional circulation delivers mm-sized grains onto the circumplanetary disk (on the order of $10^{-8}$ to $10^{-10}$ $\rm{M_{Jup}/yr}$, calculated at the Hill-sphere), bridging above the midplane. The solids are accreting in vertical direction from above the planet. Contrary to previous beliefs, the mm-sized dust can reach the planet vicinity due to the meridional circulation, and it feeds the circumplanetary disk continuously with gas and solids. To form the moons, the delivered dust can coagulate inside the circumplanetary disk \citep{DSz18} to build up planetary satellites (moons).
    \item The Hill-sphere accretion rates for the dust and gas are increasing with planetary mass, but there is no dependence on semi-major axis. 
    \item The Hill sphere gains $10^{-6}$ to $10^{-8}$ $\rm{M_{Jup}/yr}$ of gas. The difference between the dust and gas accretion rates is smaller for smaller planetary masses. The mm-grains are trapped more easily in the vicinity of the planet than the gas, which means the circumplanetary disk might be enriched with solids in comparison to the global dust-to-gas ratio (DTG) in the circumstellar disk.
    \item With each passage of the planet and its wakes, dust is stirred by the planet's spiral wakes. After the planet has passed, the dust stirred up dust settles toward the midplane (but the settling timescale is generally larger than the orbital timescale of the planet). Therefore, multiple layers of dust are created in a fan-like structure.
    \item We calculated the local dust-to-gas ratio everywhere in the circumstellar disk. It ranges from gas-only regions to values of up to 27. We calculated the surface heights where the dust-to-gas ratio is 1.0, 0.1, 0.01 and 0.001 (see Tab. \ref{tab:dtg}). These surfaces can be found between 0.2 to 7 degrees above the midplane, increasing with planetary mass. For Jupiter-mass planets and more massive giants, the streaming instability condition of DTG=1.0 can not only be found in the midplane, but even above. 
    \item The spiral wakes have different opening angles in the midplane compared to the regions above, due to the vertical temperature gradient of the disk, as it was found also by \citet{Rosotti20}, and the appearance of buoyancy spirals. This effect also influences the distribution of dust in the higher disk altitudes.
    \item Due to the vertical temperature gradient, the gap width is not the same at the different altitudes above the midplane, similarly as it was found for gas gaps in \citet{Szulagyi17}.
    \item All the listed results above turned out to be not really sensitive to the semi-major axis of the planet, they are mainly dependent on the planetary mass.
\end{itemize}

This paper is a continuation of \citet{Binkert21}, where we made ALMA synthetic images, compared gap profiles, created an equation linking the width of ALMA gaps with the planetary mass (based on semi-major axis of the planet, beam-size, wavelength etc.), and compared the disk masses from the hydrodynamic simulations with the masses derived from the ALMA mock observations. There we found that, due to the dust stirring, the ALMA disk masses are greatly underestimated.

\begin{acknowledgments}
We thank our referee, Clement Baruteau, for his great suggestions to improve our manuscript and his thoughtful review. J.Sz. thanks for the financial support through the Swiss National Science Foundation (SNSF) Ambizione grant PZ00P2\_174115. Furthermore, these results are part of a project that has received funding from the European Research Council (ERC) under the European Union’s Horizon 2020 research and innovation programme (Grant agreement No. 948467). Computations partially have been done on the "Piz Daint" machine hosted at the Swiss National Computational Centre and partially carried out on ETH Z\"urich's Euler machine. F.B. acknowledge funding from the Deutsche Forschungsgemeinschaft under Ref. no. FOR 2634/1 and under Germany's Excellence Strategy (EXC-2094–390783311).
\end{acknowledgments}

\bibliography{sample631}{}
\bibliographystyle{aasjournal}



\end{document}